# Comparative Evaluation of Community Detection Algorithms: A Topological Approach

**Günce Keziban Orman**[1,2], **Vincent Labatut**[1], **Hocine Cherifi**[2]

[1]Galatasaray University, [2]University of Burgundy

korman@gsu.edu.tr, vlabatut@gsu.edu.tr, hocine.cherifi@u-bourgogne.fr

**Abstract:** Community detection is one of the most active fields in complex networks analysis, due to its potential value in practical applications. Many works inspired by different paradigms are devoted to the development of algorithmic solutions allowing to reveal the network structure in such cohesive subgroups. Comparative studies reported in the literature usually rely on a performance measure considering the community structure as a partition (Rand Index, Normalized Mutual information, etc.). However, this type of comparison neglects the topological properties of the communities. In this article, we present a comprehensive comparative study of a representative set of community detection methods, in which we adopt both types of evaluation. Community-oriented topological measures are used to qualify the communities and evaluate their deviation from the reference structure. In order to mimic real-world systems, we use artificially generated realistic networks. It turns out there is no equivalence between both approaches: a high performance does not necessarily correspond to correct topological properties, and vice-versa. They can therefore be considered as complementary, and we recommend applying both of them in order to perform a complete and accurate assessment.

## 1. Introduction

As a modelling tool, the complex network paradigm has spread through many application fields during the last decades: biology, sociology, physics, computer science, communication, etc. (see [1] for a very complete review of applied studies). Once the model is established, the resulting network can be analysed or visualized using some of the many tools designed for graph mining. Such large real-world networks are characterized by a heterogeneous structure, leading to specific properties. In particular, a heterogeneous distribution of links often results in the presence of a so-called community structure [2]. A community roughly corresponds to a group of nodes more densely interconnected, relatively to the rest of the network [3]. The way such a structure can be interpreted is obviously dependent on the modelled system. However, independently from the nature of this system, it is clear the community structure conveys some very important information, necessary to a proper understanding [4]. Detecting communities is therefore an essential part of modern network analysis.

There are many different community detection algorithms, as reported in [2]. They can differ in two ways: not only the process leading to an estimation of the community structure, but also the nature of the estimated communities themselves. This raises a question regarding the comparison of these algorithms, from both a theoretical and a practical point of view. Authors traditionally test their community detection algorithms on real-world [5, 6] and/or artificial networks [5-7]. Their performance is classically assessed in terms of node membership, by considering the community structures as partitions and comparing the communities simply as node sets. The estimated communities are compared to some communities of reference using an association measure such as the Normalized Mutual Information [8]. The resulting single score is then compared with those obtained when applying pre-existing algorithms on the same data.

The main problem with this approach is it completely ignores the topological nature of the communities: two algorithms can reach the exact same level of performance, but still estimate community structures with very different link distributions. Thus, it seems important to take this information into account when comparing the community structures. For this purpose, we propose to use some of the community-oriented topological measures recently defined in the literature [4, 9]. Measures such as community-wise density, average distance, internal transitivity, hub dominance, and embeddedness give a detailed description of the topology of the communities and their interactions. Up to now, they have been used to characterize and compare the community structures of real-world networks, but never to evaluate community detection algorithms. We consider them as an alternative or an addition to the traditional performance measures mentioned earlier. The question is then to know whether this new topological assessment leads to the same results than the partition-based approach.

Besides the method used to evaluate the algorithms, another important point concerns the data used during this evaluation. It is well known that, when solving a search problem, no algorithm can be superior on all possible instances of the problem [10]. It is therefore necessary to focus on a specific subset when comparing algorithms. Community detection algorithms are usually designed to study real-world systems, so it is natural to consider only the networks representing them. However, using only real-world networks is an issue because the identification of their community structure implies an expert human intervention, making them rare and/or relatively small. Artificial networks constitute an appealing alternative, because they can easily be generated in large amounts. All that is needed is a generative model able to produce networks with realistic topological properties. Some properties common to most real-world networks are well-identified: power-law distributed degree, small-worldness, non-zero degree correlation and relatively high transitivity [11]. Additionally, networks with a community structure are characterized by a power-law distributed community size [12]. Several generative models with increasing realism were successively designed [7] before finally meeting these constraints [7, 13, 14].

In this article, we use the LFR model [7] with appropriate parameters to generate realistic undirected and unweighted networks. We study their topological properties to assess their realism. We then apply a representative selection of community detection algorithms to these networks. We evaluate the quality of the estimated community structures using both the classic performance measures and the topological approach, and we compare the obtained results. In section 2, we review the performance measures traditionally used when comparing community detection algorithms and we then describe the properties we selected to characterize the topology of community structures in section 3. In section 4, we review the various approaches used in the literature to define the concept of community, and select a representative set of community detection algorithms from this perspective. In section 5, we describe the LFR model we used to generate the artificial networks constituting our benchmark. Additionally, we introduce an adjustment allowing to improve their realism. In section 6, we first analyse the topological properties of the generated community structures. We then focus on the evaluation of the algorithms, both from a traditional and topological point of view. Finally, we discuss our results and explain how our work could be extended.

## 2. Performance of Community Detection Algorithms

The traditional methods used to assess the performance of community detection algorithms consider a community structure as a partition of the node set. Comparing the community structure estimated by the algorithm with the reference community structure therefore consists in comparing two partitions (estimated and reference). Many measures exist for this purpose, and this problem is very classic, so we briefly describe the most widespread ones here.

The *Fraction of Correctly Classified nodes* (FCC) has been used by several authors, the first seeming to be Girvan and Newman [6]. According to this measure, a node is correctly classified if its estimated community is the same than for the majority of nodes present in its reference community. Moreover, if an estimated community corresponds to a fusion of several reference communities, all the concerned nodes are considered as misclassified [15, 16]. The total number of correctly classified nodes is divided by $n$ to normalize the measure, which results in a value between $0$ and $1$.

The *Rand Index* (RI) [17] corresponds to the proportion of node pairs for which both the estimated and reference community structures agree. For a given pair, there is agreement when both nodes belong to the same community, or to different communities, for both community structures. Consequently, there is disagreement if the nodes are in the same community for one community structure, whereas they belong to two different ones for the other. The Rand Index ranges from $0$ (the algorithm completely failed to estimate the community structure), to $1$ (the algorithm perfectly estimated the community structure). The *Adjusted Rand Index* (ARI) is a corrected for chance version of the RI [18] ranging from $-1$ (less than chance agreement) to $1$ (complete agreement). Zero corresponds to a pure chance agreement.

The *Normalized Mutual Information* measure (NMI) was defined in the context of classical clustering [19] to compare two different partitions of one data set, by measuring how much information they have in common. Danon *et al.* [16] used it to assess the performance of community detection algorithms, and the measure was subsequently used by various other authors [7]. If the estimated communities correspond perfectly to the reference ones, the measure takes the value $1$, whereas it is $0$ when they are independent.

## 3. Community-Oriented Topological Properties

The performance measures presented in the previous section allow evaluating the similarity between two community structures, by considering them as partitions. They are compared only in terms of individual node membership, without taking links into account. In this section, we present some topological properties defined to characterize community structures. Unlike the previous measures, they can be used to compare two partitions from a purely topological perspective. In this sense, they allow to evaluate the quality of an algorithm in a complementary way. Using these measures, a recent study showed the real-world networks possessing a community structure display certain differences [4], which contrasts with what we mentioned in the introduction regarding the existence of topological properties common to most real-world networks. These differences appear to be characteristic of the considered real-world system, which allows the authors to identify five different classes: Communication, Internet, Information, Biological and Social networks. In this section, we describe the main properties and give their typical real-world values for those classes.

### 3.1. Embeddedness

The *embeddedness* measure assesses how much the direct neighbours of a node belong to its own community. It is defined as the ratio of the internal degree $k_{int}$ to the total degree $k$ of the considered node [4]:

$$e = k_{int}/k \tag{1}$$

This *internal degree* is the number of links the node has with other nodes from the same community, by opposition to its *external degree* $k_{ext}$, which corresponds to connections with nodes located in other communities. The maximal embeddedness of 1 is reached when all the neighbours are in its community ($k_{int} = k$) whereas the minimal value of $0$ corresponds to the case where all neighbours belong to different communities ($k_{int} = 0$). In real-world networks, a majority of nodes, usually with low degree, have a very high embeddedness (so almost no links outside their community). For the remaining nodes, the embeddedness distribution depends on the considered class. Communication, Internet and biological networks exhibit a peak around $e = 0.5$, whereas social and information networks have a more uniform distribution. In all cases, the whole range of $e$ is significantly represented, even small values [4].

## 3.2. Community Size

The *community size distribution* is considered as an important characteristic of the community structure. It has been largely studied in real-world networks, and seems to follow a power-law [12] with exponent $\beta$ ranging from 1 to 2 [5]. This means the community sizes are heterogeneous, with many small communities and only a few very large ones. In real-world networks, the minimal community size is 2, but, the maximal community size varies widely depending on the class and the granularity of the modelled system [4].

## 3.3. Internal Transitivity

The *internal transitivity* is based on the classic local transitivity, averaged over the nodes located inside a community. The *local transitivity* of a given node depends on how its direct neighbours are interconnected. It is defined as the number of links present between these neighbours, divided by the number of links one would get if they were all interconnected. In other words, it represents the proportion of existing to possible links in the node neighbourhood. The internal transitivity for some community $C$ is formally defined as:

$$T(C) = \frac{1}{n_C} \sum_{i \in C} \frac{2 \times l(i)}{k_{int}(i)[k_{int}(i) - 1]} \qquad (2)$$

Here, $n_C$ is the number of nodes in community $C$, $l(i)$ is the number of links among the neighbours of some node $i$ which all belong to the same community, and $k_{int}(i)$ is the internal degree of some node $i$ (as defined previously for the embeddedness). In real-world networks, the distribution of internal transitivity varies with the community size in different ways, as shown in [4]. It increases for Internet and communication networks. For biological and social networks, it increases until it reaches a peak value, and then starts decreasing.

## 3.4. Scaled Density

The density $\rho$ of a community $C$ is defined as the ratio of links it actually contains, noted $m_C$, to the number of links it could contain if all its nodes were connected. In the case of an undirected network, the latter is $n_C(n_C - 1)/2$, where $n_C$ is the number of nodes in the community, and we therefore get $\rho = 2m_C/(n_C(n_C - 1))$. When compared to the overall network density, the community density allows assessing the cohesion of the community: by definition, a community is supposed to be denser than the network it belongs to. The *scaled density* is a variant obtained by multiplying the density by the community size [4]:

$$\tilde{\rho}(C) = \rho(C) n_c = 2m/(n - 1) \qquad (3)$$

If the considered community is a tree, it has only $m_C = n_C - 1$ links, and $\tilde{\rho}(C) = 2$. If it is a clique (completely connected subnetwork), then $m_C = n_C(n_C - 1)/2$ and we have $\tilde{\rho}(C) = n_C$. The scaled density therefore allows characterizing the structure of the community. Some real-world networks such as the Internet or communication networks have essentially tree-like communities. On the contrary, for other classes like social and information networks, the scaled density increases with the community size. Finally, biological networks exhibit a hybrid behaviour, their small communities being tree-like whereas the large ones are denser and close to cliques [4].

## 3.5. Average Distance

The *distance* between two nodes corresponds to the length of their shortest path. When averaged over all pairs of nodes in a community, it allows assessing the cohesion of this community. In real-world networks, small communities ($n_C \leq 10$) are supposedly small-world, which means that, over the whole network, the community average distance $\ell$ should increase logarithmically with the community size

$n_C$ [4]. For larger communities, the average distance still increases, but more slowly or even stabilizes for certain classes of real-world networks like communication networks. A small average distance can be explained by a high density (social), the presence of hubs (communication, Internet), or both (biological, information).

*3.6. Hub Dominance*

From a community structure perspective, a hub is a node connected to many of the other nodes belonging to the same community. The presence of a central hub in a community $C$ can be assessed using the *hub dominance* measure, which corresponds to the following ratio:

$$h(C) = \max_{C}(k_{int})/(n_C - 1) \qquad (4)$$

The numerator is the maximal internal degree found in $C$, and the denominator is the maximal degree theoretically possible given the community size. The hub dominance therefore reaches 1 when at least one node is connected to all other nodes in the community. It can be 0 only if no nodes are connected, which is unlikely for a community. In real-world networks, the behaviour of this property depends on the considered class. For communication networks, it is close to the maximum for all community sizes, meaning hubs are present in all communities. Considering their communities are sparse and tree-like, one can conclude they are star-shaped. Other classes do not have as many hubs in their large communities, which is why their hub dominance generally decreases with community size increase [4].

## 4. Community Detection Algorithms

A very widespread informal definition of the community concept considers it as a group of nodes densely interconnected compared to the other nodes [2, 20]. In other words, a community is a cohesive subset clearly separated from the rest of the network. However, formal definitions differ in the way they translate and combine both these aspects of cohesion and separation. In the literature, there are numerous community detection algorithms, many implementing differently the notion of community structure. Here, we selected a representative set of algorithms and categorized them according to the method they apply to identify communities. We chose to ignore some algorithms because they were too slow to be included in this study (e.g. node-removal approaches such as Edge-Betweenness [6]).

*4.1. Modularity-Based Approaches*

A direct translation of the informal definition given above consists in first specifying two distinct measures to assess separately cohesion and separation, and then processing an overall measure by considering their difference or ratio. This approach led to many variants, differing on how the measures are defined and combined. The most widespread one is certainly the modularity, a chance-corrected measure which assesses cohesion and separation through the number of intra- and inter-community links, respectively [21]. We selected two modularity optimization algorithms, which differ in the way they perform this optimization.

**Fast Greedy** applies a basic greedy approach [15]. It starts with a state in which each node is in its own community and the algorithm repeatedly joins pairs of communities together to obtain larger ones. At each step, the joined communities are selected by considering the largest increase (or smallest decrease) in modularity. By definition of the modularity, communities connected by many links will be favoured. Thanks to this agglomerative approach, FastGreedy produces a set of community structures organized hierarchically, with increasing granularity. The one obtaining the maximal modularity is considered as the best. This algorithm is relatively fast, however, the result of the greedy optimization can be very coarse.

**Louvain** also adopts an agglomerative hierarchical method, but it relies on a slightly different greedy optimization process, and includes an additional aggregation step to improve processing on large networks [22]. Like for FastGreedy, each node is initially placed in its own community. For each node, the modularity gains obtained by moving it into each one of its neighbours' community is calculated. The node is then moved in the community associated to the largest gain, or stays in its original community if no gain is possible. Louvain applies this procedure repeatedly and sequentially for all nodes until no further improvement can be achieved, which leads to the end of this first step. The second step consists in building a new network whose nodes are the communities estimated during the first step. The inter- and intra-community links are represented in the new network by weighted regular links and self-loops, respectively. The first step is then applied to this network, and both steps are repeated until stable communities are reached.

*4.2. Node Similarity-Based Approaches*

Another category of approaches is based on node similarity measures. Such a measure allows translating the topological notions of cohesion and separation in terms of intra-community similarity and inter-community dissimilarity. In other words: a community is viewed as a group of nodes which are similar to each other, but dissimilar from the rest of the network. Once all node-to-node similarities are known, detecting a community structure can be performed by applying a similarity-based classic cluster analysis algorithm [23]. In this work, we have used one algorithm from this class.

***WalkTrap*** uses a distance measure based on random walks and applies a hierarchical agglomerative clustering approach [24]. A random walker is an agent moving from one node to another following the network links. At each time step, the next node is selected by randomly picking a neighbour of the current node. The idea behind this algorithm is that random walks tend to get trapped into a community, hence its name. If two nodes $i$ and $j$ are in the same community, the probability to get to a third node $k$ located in the same community through a random walk should not be very different for $i$ and $j$. The distance is constructed by summing these differences over all nodes, with a correction for degree.

*4.3. Compression-Based Approaches*

Some approaches based on data compression do not use the cohesion and separation concepts like the previous definitions. They consider the community structure as a set of regularities in the network topology, which can be used to represent the whole network in a more compact way than the whole adjacency matrix. The best community structure is supposed to be the one maximizing compactness while minimizing information loss. The quality of the representation is assessed through measures derived from information theory. Algorithms essentially differ in the way they represent the community structure and how they assess the quality of this representation. We have used two algorithms from this class.

***InfoMod*** uses a simplified representation of the network focusing on the community structure: a community matrix and a membership vector [25]. The former is an adjacency matrix defined at the level of the communities (instead of the nodes), which means its size is $k \times k$. The latter is a vector of size $n$, associating each node to a community. The amount of information from the original network contained in the simplified representation is quantified using the mutual information measure. Among all possible assignment of nodes to communities, the best is the one associated with the maximal mutual information. This optimisation is performed by simulated annealing. The selection of the optimal number of communities is realized using the minimum description length principle.

***InfoMap*** represents the community structure through a two-level nomenclature based on Huffman coding [26]: one level to distinguish communities in the network and the other to distinguish nodes in a community. The problem of finding the best community structure is expressed as minimizing the quantity of information needed to represent some random walk in the network using this nomenclature. With a partition containing few inter-community links, the walker will probably stay longer inside communities, therefore only the second level will be needed to describe its path, leading to a compact representation. The authors optimize their criterion using simulated annealing.

*4.4. Significance-Based Approaches*

A completely different approach consists in considering the statistical significance of the whole community structure, or of the individual communities. A community structure can be expected under certain circumstances, but groups of densely connected nodes can also appear only by chance. For instance, even a generative model not supposed to create any community structure can produce clusters of nodes due to random fluctuations. Statistical significance allows distinguish them from actual communities. We use one algorithm from this category.

*Order Statistics Local Optimization Method* (OSLOM) is a local optimization method applied to a score measuring the statistical significance of individual communities [27]. The authors define this statistical significance as the probability of finding a similar community (same size, degree sequence and internal connections) in a null model possessing no community structure. In its first phase, OSLOM agglomerates neighbour nodes to obtain a collection of significant, possibly overlapping communities. Adjustments are performed by removing/adding nodes to communities, in order to increase their significance. Due to its stochastic nature, this process is repeated several times to ensure stability. Various hierarchical levels are obtained by applying recursively the same principle to the resulting supernetwork (network whose nodes represent communities in the lower hierarchical level). OSLOM allows detecting mutually-exclusive, overlapped or hierarchical communities in simple, directed or weighted networks. However, as previously mentioned, here we use it only on undirected unweighted networks.

*4.5. Diffusion-Based Approaches*

Diffusion-based approaches tackle the problem of community detection using a communication paradigm. They rely on the assumption that information is more efficiently exchanged between nodes of the same community. Therefore, communities can be detected by considering how information is propagated in the network.

*Community Overlap Propagation Algorithm* (COPRA) is an extended version of Raghavan *et al.*'s *Label Propagation* algorithm [28], proposed by Gregory [29]. The information takes the form of a label, and the propagation mechanism relies on a vote between neighbours. Initially, each node is labelled with a unique value. Then an iterative process takes place, where each node takes the label which is the most spread in its neighbourhood (ties are broken randomly). This process goes on until convergence, i.e. each node has the majority label of its neighbours. Communities are then obtained by considering groups of nodes with the same label. By construction, one node has more neighbours in its community than in the others. This algorithm is faster than most other algorithms. We apply Gregory's version to detect mutually exclusive communities in undirected unweighted unipartite networks. However, note it is able to handle overlapping communities, for both weighted and bipartite networks.

*MarkovCluster* simulates a diffusion process in the network to detect communities [30]. This approach relies on the transfer matrix, which describes the transition probabilities for a random walker evolving in this network. Two transformations, expansion and inflation, are iteratively applied to this matrix until convergence. The expansion operation raises the transfer matrix to a power $p$. The result is a matrix showing the probability that a random walker starts from node $i$ and reaches node $j$ in $p$ steps. The inflation operation consists in raising each element in the matrix to some specified power, in order to favour the higher probability values. These correspond to pairs of nodes presumably belonging to the same community. The value of this power has a direct effect on the granularity of final communities. The resulting matrix is then normalized to get a new transfer matrix, and the process is repeated until convergence. The final matrix can be interpreted as the adjacency matrix of a network with disconnected components, which correspond to communities in the original network.

## 5. Generative Model

We selected the LFR model to generate artificial networks with a community structure. By construction, this model guaranties to obtain values considered as realistic [1, 11] for several properties: size of the network, power-law distributed degrees and community sizes. The model allows to control directly the following properties: number of nodes $n$, desired average $\langle k \rangle$ and maximal degrees $k_{max}$, exponent $\gamma$ for the degree distribution, exponent $\beta$ for the community size distribution, and mixing coefficient $\mu$. The latter represents the desired average proportion of links between a node and nodes located outside its community, called inter-community links. Consequently, the proportion of intra-community links is $1-\mu$. A node of degree $k$ has therefore an external degree of $k_{ext} = \mu k$ and an internal degree of $k_{int} = (1-\mu)k$. By definition, it is rather clear the mixing coefficient is complementary to the embeddedness presented in section 3: $e = 1 - \mu$.

The generative process first uses the Configuration model [31] to generate a network with average degree $\langle k \rangle$, maximal degree $k_{max}$ and power-law degree distribution with exponent $\gamma$. Second, virtual communities are defined so that their sizes follow a power-law distribution with exponent $\beta$. Each node is randomly affected to a community, provided the community size is greater or equal to the node internal degree. Third, an iterative process takes place to rewire certain links, in order to approximate $\mu$, while preserving the degree distribution. For each node, the total degree is not modified, but the ratio of internal and external links is changed so that the resulting proportion gets close to $\mu$. The main network properties have been studied empirically on this model [14]. It turns out LFR generates small-world networks, with relatively high transitivity and degree correlation.

In order to select appropriate values for the network parameters, we considered several studies of real-world networks [4, 11]. For the power-law exponents, we used $\gamma = 3$ and $\beta = 2$, which seem to be the most representative values. Concerning the numbers of nodes and links, however no typical values emerge. Those studies show real-world complex networks can have very different sizes, defined on a wide range going from tens to millions of nodes. The average and maximal degrees are also very variable, so it is difficult to characterize them, too. As a result, we selected some consensual values for these parameters, while considering also the computational aspect of community detection.

Table 1. Properties of the real-world and generated artificial networks. $n$ is the network size, $\langle k \rangle$ and $k_{max}$ its average and maximal degree, $n_C^{min}$ and $n_C^{max}$ the sizes of its smallest and largest communities.

| Origin | Name | $n$ | $\langle k \rangle$ | $k_{max}$ | $n_c^{min}$ | $n_c^{max}$ |
|---|---|---|---|---|---|---|
| *Real-world* | *dmela* | 7500 | 6 | 180 | 2 | 200 |
| | *caida* | 25000 | 6 | 3000 | 2 | 3000 |
| | *email* | 250000 | 2 | 10000 | 2 | 10000 |
| *Generated* | *#1* | 7500 | 10 | 180 | 2 | 180 |
| | *#2* | 25000 | 11 | 2850 | 3 | 2300 |
| | *#3* | 250000 | 5 | 7275 | 2 | 8115 |

In the LFR model, the embeddedness depends on a single parameter, the mixing coefficient $\mu$, which controls the level of separation of the community structure. By construction, the embeddedness distribution is very strongly peaked near the value $1-\mu$. Yet, as we mentioned previously, in real-world networks the embeddedness has a different distribution, which depends on the considered network class. To overcome this drawback, we modified the LFR model so that it produces a more realistic embeddedness distribution [32]. After some tests, we decided to focus on three classes in particular, because the generated networks were globally more similar to them: communication, Internet and biological networks. In our modified model, the obtained mixing coefficient $\mu$ is distributed accordingly: normally (over its whole definition domain $[0;1]$) for half the nodes, whereas it is $0$ for the other half (nodes connected only to other nodes from the same community).

The upper part of Table 1 summarizes the main properties measured for some networks representing the three classes we wanted to match: biological (*dmela*), Internet (*caida*) and communication (*email*) networks. The lower part focuses on the networks we generated, which display very close properties. These are the most realistic networks we can get with respect to the current knowledge. We generated samples of 5 networks for each network size, in order to insure consistency.

## 6. Results

We analysed the properties of the generated networks and communities to check their realism level. After this, we applied the eight selected community detection algorithms from section 4 and assessed their results, not only from a community membership point of view as explained in section 2, but also by considering the topological properties described in section 3.

*6.1. Topological Properties of the Generated Networks and Communities*

Among the community-related properties we described in section 3, two are directly controlled by the LFR model: the community size and embeddedness distributions. Our measurements confirm on all networks that the community sizes follow a power-law distribution as expected (cf. Figure 1). Note the range of these sizes varies much from one real-world network to the other, and it is therefore difficult to describe a typical set of values. However, as we have indicated in Table 1, the minimal and maximal community sizes are very similar to those observed in real-world networks of comparable size. For the embeddedness, we obtained 1 for half the nodes, as expected. But for the remaining half, although we used a normal distribution during the generation process, we obtain more nodes with a 0 embeddedness than expected. We assume that this result is due to some incompatibilities between the mixing coefficient and degree affected to certain nodes. For instance, a node whose degree is only 1 cannot have a 0.2 embeddedness: the LFR rewiring process will certainly give it a 0 embeddedness. We plan to correct this point in our future work. Those results are nonetheless close enough to what can be observed in biological, information and communication networks.

We now focus our attention on the uncontrolled properties. The scaled density increases from 2 to 14 for $n=7500$ and 25000 and to 8 for $n=250000$, along with the community size. This means the smallest communities are tree-like ($\tilde{\rho}(C)=2$) which is consistent with what is observed in real-world networks. However, for larger communities ($n_C > 5$), there is no tree-like or clique-like ($\tilde{\rho}(C)=n_C$), structures, which is not realistic. It seems the links are distributed more homogeneously over the generated networks, making small communities too dense and large ones too sparse.

As shown in Figure 1 the average distance increases regularly from 1 to 2 for $n=7500$, 1.5 to 2.5 for $n=25000$ and 1.5 to 3 for $n=250000$ until a community size limit (approximately 20, 35 and 100 respectively), and then stays stable. For $n=25000$ and 250000 in particular, we observe a fast increase: this is characteristic of real-world networks. For the rest of the communities, the observed distribution is also comparable to communication networks, with a stable average distance for medium and large communities, though. Moreover, the values measured for these communities are also realistic in terms of magnitude. This average distance distribution is especially similar to the biological, information and communication networks that we used as references.

We observe that hub dominance decreases for small communities along with community size increase, but for medium and large communities it starts to increase again. Its trend is similar for all network sizes, but the peak values and community sizes triggering the increase and decrease are different. For $n=7500$, hub dominance is very high for the smallest communities, with values close to 1. Then it decreases until 0.8 with community size increase and it stays relatively stable after this value. For $n=25000$ and 250000, the decrease and the following increase are more evident. The hub dominance value for small communities is around 0.8 for both sizes. It decreases very much until 0.4 for $n=250000$ when community size is around 20 and 0.6 or $n=25000$ for community size 10.

For larger communities, we also observe a higher dispersion between the different iterations. The same dispersion was also observed on real-world networks [4]. Hub dominance completely relies on the way high degree nodes are distributed over communities, since it directly depends on the maximal internal degree found in communities. The fact there are much less large communities, due to their power law-distributed sizes, can explain this dispersion. A possible solution would be to consider a measure based on the $k$ highest internal degrees of the community instead of a single one.

The internal transitivity decreases when the community size increases. This decrease is more evident for $n=7500$, ranging from 0.8 to 0.2. But for larger networks, we observe neither a high transitivity for small communities, nor a clear decrease. Especially, for $n=250000$ we obtain very low values. In real-world networks, the internal transitivity undergoes different trends: either an increase along community size or an initial increase followed by a decrease. None are consistent with what we observe on the generated networks. The LFR model relies on the Configuration model to generate its initial scale-free network, and this model is known to produce networks with a very low transitivity. As we showed in a previous study [13], the relatively high overall transitivity observed in the final network is due to the rewiring process of LFR. From the results presented here, we can conclude this effect seems to be stronger for the small communities than the large ones. This can be explained by the fact the rewiring does not aim at forming triangles: it randomly selects nodes in the same community. Therefore, it is more probable to inadvertently create triangles if the community is small.

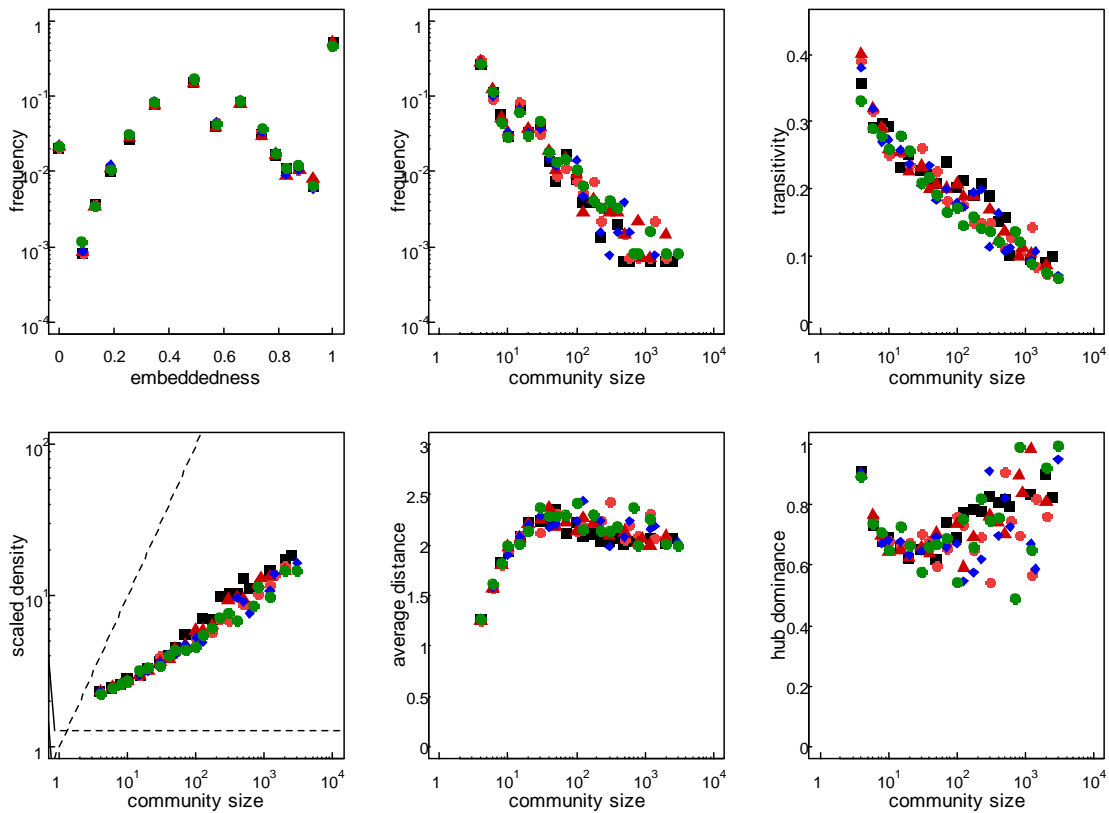

Figure 1. Properties of the generated communities for $n=25000$, $\gamma=3$ and $\beta=2$. Each one of the 5 networks in the sample is represented with a different shape/colour. Points represent averages over logarithmic bins of the community size.

To summarize our observations: the generated communities exhibit most of the properties observed in real-world networks. The measured values vary according to the network size, but the trends are very similar. The most realistic results have been obtained for $n = 25000$. The distributions of community sizes, average distance, hub dominance and scaled density, especially for small communities, are globally realistic. However, this is not as true for the internal transitivity. Another issue is the fact the generated networks do not comply with a specific class of real-world networks, but rather have similarities with different classes depending on the considered property: the average distance distribution is reminiscent of communication networks, but embeddedness and hub dominance distributions are more related to social and biological networks. However, despite those limitations, the modified LFR model allows us to generate the most realistic artificial networks to date.

*6.2. Traditional Evaluation of the Algorithms*

We applied the selected community detection algorithms on the generated networks. It is worth recalling our goal is not to actually assess the performance of these algorithms, but rather to discuss the evaluation methods themselves. For this reason, each algorithm is simply used with its default parameters. Moreover, COPRA and OSLOM are forced to detect mutually exclusive communities (i.e. no overlap allowed). We focused on the networks of size $n = 25000$, which are the most realistic, according to our analysis. We then processed the algorithms performances using all the measures presented in section 2. The results are listed in Table 2.

Table 2. Traditional performance measure values and their ranking for all eight algorithms

| Algorithm | FCC Value | Rank | RI Value | Rank | ARI Value | Rank | NMI Value | Rank |
|---|---|---|---|---|---|---|---|---|
| COPRA | 0.090 | 7 | 0.068 | 8 | 0.002 | 8 | 0.070 | 8 |
| FastGreedy | 0.080 | 8 | 0.919 | 7 | 0.272 | 6 | 0.588 | 7 |
| InfoMap | 0.862 | 2 | 0.997 | 1 | 0.930 | 1 | 0.930 | 1 |
| InfoMod | 0.255 | 6 | 0.971 | 5 | 0.256 | 7 | 0.620 | 6 |
| Louvain | 0.425 | 4 | 0.982 | 3 | 0.692 | 3 | 0.735 | 4 |
| MarkovCluster | 0.881 | 1 | 0.993 | 2 | 0.822 | 2 | 0.881 | 2 |
| OSLOM | 0.415 | 5 | 0.932 | 6 | 0.337 | 5 | 0.685 | 5 |
| WalkTrap | 0.818 | 3 | 0.979 | 4 | 0.614 | 4 | 0.865 | 3 |

All measures, except FCC, agree on the fact InfoMap finds the community structure the most similar to the reference. For FCC, MarkovCluster is ranked first, whereas for the other measures, it is the second best algorithm. WalkTrap and Louvain are the third or fourth best algorithms according to all results. However, if the obtained values are very close for the measures ranking Louvain third (RI and ARI), the difference is much more important when WalkTrap is third (FCC and NMI).

OSLOM is following Louvain as it is ranked fifth by all measures except RI, which puts it in sixth position. InfoMod and FastGreedy follow, and for them the difference between the measures in terms of amplitude of the measured performance is much more important than for the other algorithms. Especially, for FCC, we observe a large difference between FastGreedy ($0.08$) and the other algorithms ($> 0.25$). COPRA is the last algorithm according to RI, ARI and NMI while it is seventh for FCC. Its performances are close to zero and clearly lower than all other algorithms especially for RI, ARI and NMI. This might be due to the fact it was initially designed to detect overlapping communities, which is not the case here.

For several algorithms, the values obtained with FCC are much lower than for the other measures: FastGreedy, Louvain, InfoMod, COPRA and OSLOM. As explained in section 2, when an estimated community corresponds to several reference ones, FCC considers all nodes as misclassified. So we can suppose these algorithms tend to detect large communities corresponding to the merge of several actual ones. We do not observe high differences between the algorithms performances for RI. On the contrary, for ARI, which is the chance-corrected version of RI, we see clear differences, especially for FastGreedy, OSLOM, COPRA and InfoMod. Among all measures, FCC and ARI give the most contrasted results. However, in terms of algorithms ranking, all measures agree with very small divergences. Let us consider the algorithms performance in function of the algorithms categories we defined in section 4. The performance of InfoMap and InfoMod, which belong to the compression-based approach category, are very different. Similarly, the performance of FastGreedy and Louvain, both belonging to the modularity optimization category, are quite different. This seems to show that, although some algorithms use the same idea to define what a community is, the process they implement to perform the partitioning affects significantly their performance.

*6.3. Topological Evaluation of the Algorithms*

The topological properties of the algorithm-estimated and reference structures are shown on Figure 2. The black line shows the properties of the reference structure and each shape/colour corresponds to one algorithm. In order to verify if the sizes of the estimated communities follow a power-law like the reference, we applied a test of goodness of fit. As shown on Figure 2, it is the case for WalkTrap, MarkovCluster and InfoMap, with p-values of $0.33$, $0.31$ and $0.12$, respectively. However, the other algorithms (FastGreedy, Louvain, OSLOM, COPRA and InfoMod) find community structures whose size distributions are significantly different from a power-law ($p$-value smaller than $0.001$). This can be explained by a thorough examination of the identified communities. The smallest community that InfoMod identified has around $20$ nodes, while it is $3$ for the reference. The largest community found by FastGreedy is close to $10000$, which is much larger than the largest reference community size, around $2000$. COPRA puts almost all nodes into one community and the rest of the nodes are put in their own community, or very small ones. Thus its largest community contains more than $20000$ nodes. These remarks are consistent with the assumption we made to explain the very low FCC values obtained in the previous subsection, regarding the possibility that many communities identified by these algorithms may correspond to merged reference ones. To a lesser extent, MarkovCluster and WalkTrap also find numerous single-node communities, but this does not seem to affect traditional performance measures much. Note the inflation parameter of MarkovCluster determines the desired granularity of the communities. We use the default value; it is likely larger communities would be found if using higher values.

For the embeddedness, WalkTrap, FastGreedy, Louvain, OSLOM and InfoMap are very close to the reference values. The frequency of the nodes having a zero embeddedness for InfoMap and OSLOM is lower than in the reference, though. On the contrary, the embeddedness obtained for InfoMod, COPRA and MarkovCluster is clearly different. InfoMod and MarkovCluster display more uniform distributions. For InfoMod, in particular, almost half the nodes have very low embeddedness. According to the observed embeddedness and community size distributions, we can suppose it puts in the same community many nodes which do not have links between them, and are actually in different communities in the reference structure. Thus, it finds communities larger than $20$ nodes while in reality the smallest communities contain only $3$ nodes. Because COPRA puts most of the nodes into one community, there is a large proportion of nodes with a maximal embeddedness, and the lower embeddedness values are therefore less represented.

When considering the scaled density (Figure 2), InfoMap, MarkovCluster and WalkTrap are very close to the reference, with InfoMap and WalkTrap diverging for large communities, though. For FastGreedy, OSLOM and Louvain, the scaled density is relatively stable, and does not present the slow increase which is characteristic of the reference. This can be interpreted as the fact the communities detected by these algorithms all present the same structure, independently from their size. For COPRA, the distribution corresponds to the reference for small communities ($<30$). However, its giant community does not match the reference trend; and as mentioned before, middle-sized communities are not represented at all. We observe the most different trend for InfoMod, whose scaled density decreases with the community size. This is completely opposed to what is observed for the reference. The low scaled density obtained for large communities may be the result of InfoMod putting small separated (i.e. without any connection) reference communities together in the same estimated community.

The average distances measured on the FastGreedy, Louvain, OSLOM and InfoMod communities are much dispersed and do not follow the evolution observed for the reference. FastGreedy, in particular, has a much higher average distance than the reference and the other algorithms. InfoMod has also larger average distances for larger communities. COPRA, for smaller communities, shows a trend similar to the reference, but its giant community is clearly apart, again. This property is a good indicator of cohesion, so it seems this quality is absent from the communities identified by FastGreedy, Louvain, OSLOM and InfoMod. The remaining algorithms (InfoMap, MarkovCluster, WalkTrap) are very close to the reference. The average distance of InfoMap is higher than the reference for community sizes between $10$ and $20$, though.

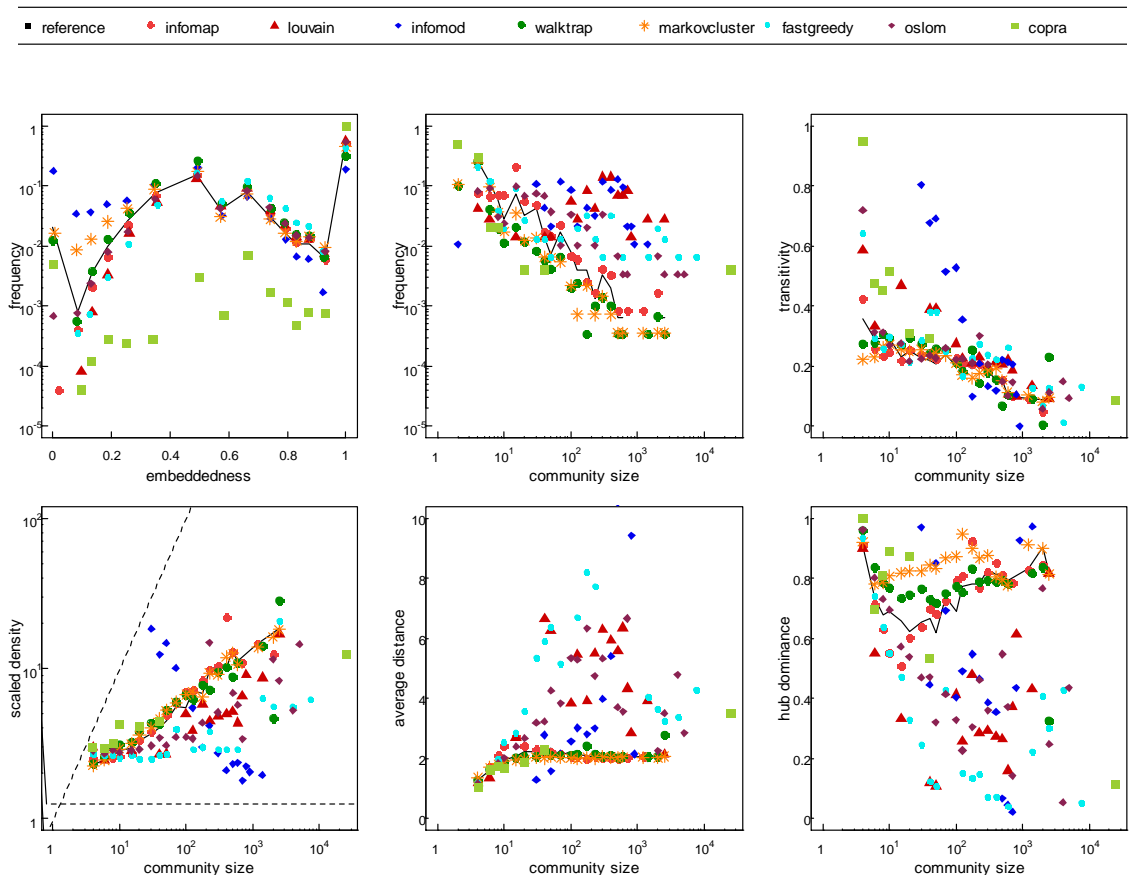

Figure 2. Properties of the detected communities for $n = 25000$, $\gamma = 3$ and $\beta = 2$. Each shape/colour corresponds to a different algorithm, whereas the reference is represented by a solid line. Points represent averages over logarithmic bins of the community size.

Unlike for the other properties, we do not observe a very high similarity between the reference and the algorithms for the hub dominance. WalkTrap and InfoMap have the most similar results. But WalkTrap hub dominance is larger than in the reference for the communities whose sizes range from $10$ to $100$, while with InfoMap it is smaller for the same communities. Moreover, the former presents an outlier for the largest community: its hub dominance is very low and does not respect the general trend. In other words, the largest communities detected by WalkTrap are much less centralized than in the reference. The hub dominance trend for MarkovCluster is also similar to the reference, but the values are higher for most of the community sizes. FastGreedy, Louvain, OSLOM, COPRA and InfoMod once again display very different behaviours compared to the reference, with hub dominance values bellow the reference ones. It is interesting to notice that for these algorithms, the hub dominance evolves in opposition with the average distance. For instance, with FastGreedy, the smallest and largest communities are centralized and have a small average distance, whereas it is the contrary for the middle-sized communities. This seems to indicate the average distance depends on the centralisation of the communities.

The general behaviour of the internal transitivity is similar to the reference for all algorithms: a decrease when the community size increases. It is especially the case for WalkTrap, OSLOM, MarkovCluster and InfoMap, with rare outliers. The values for FastGreedy and Louvain are higher than the reference for all community sizes. The small communities ($<10$) identified by COPRA display a very high transitivity. For InfoMod, the decrease is much more sudden than in the reference, reflecting the fact small communities have a very high transitivity, whereas it is very low for large ones.

We occasionally observe a similar behaviour for algorithms belonging to the same category. For instance, Louvain and FastGreedy, both based on modularity optimization, comparably differ from the reference. However, most of the time there seems to be no relationships between the category and the results. The very same Louvain and FastGreedy have very different results in terms of classic performance. According to both topological properties and performance measures, InfoMap and InfoMod, which belong to the same category, are completely opposed: they are the best and worst algorithms, respectively. Another interesting observation regarding the algorithms category concerns the similarity of the results obtained by OSLOM when compared to FastGreedy and Louvain. These algorithms directly optimize two very different criteria: statistical significance for the former, modularity for both others. They nevertheless lead to relatively similar results, not only in terms of partition comparison, but also when considering the topological properties. It therefore seems both approaches implicitly define the notion of community in a very similar way.

## 7. Conclusion

In this study, we took advantage of recent advances relative to the characterization of community structures in complex networks to propose a new, topology-based way of evaluating community detection algorithms. We first applied a variant of the LFR model [7] to generate artificial networks with known community structures. We studied their topological properties and concluded most of them are relatively realistic, which is a necessary condition for our purpose. We applied a representative set of eight fast community detection algorithms to these networks: Fast Greedy, Louvain, WalkTrap, InfoMap, InfoMod, Order Statistics Local Optimization Method (OSLOM), Community Overlap Propagation Algorithm (COPRA) and MarkovCluster. We assessed their results using first the traditional approach, which consists in evaluating the community structure quality in terms of partition comparison. It turns out the considered measures (Fraction of Correctly Classified Nodes, Rand Index, Adjusted Rand Index and Normalized Mutual Information) agree with each other with small differences when considering the way they rank algorithms: InfoMap and MarkovCluster are generally the first, followed by WalkTrap and Louvain, then OSLOM, InfoMod and FastGreedy, and finally COPRA.

We then considered the topological properties of the estimated community structures. For the extreme (first and last) algorithms, this analysis confirms the conclusions obtained with the traditional measures: InfoMap results are the most similar to the reference, whereas those of COPRA, InfoMod and FastGreedy are the most different. However, contrary to what these observations suggest, performances and topological properties do not always agree for the other algorithms. According to their performances, MarkovCluster and WalkTrap are almost as good as InfoMap, and ranked $2^{nd}$ and $3^{rd}$, respectively. But their topological features are much less similar to the reference structure, especially when considering their community size and embeddedness distributions. This is an essential point, because it means even if the performance measured for these algorithms was relatively high, the communities they identified substantially differ from the reference, topologically speaking. On the contrary, some of the properties displayed by OSLOM are relatively close to the reference, but its performance on this benchmark is far from being conclusive. So, it seems there is no equivalence between obtaining a high performance and identifying a community structure with correct topological properties. We see two reasons for that. First, each partition-based measure penalizes departures from the real community structures in a different way. For instance, some are more affected by the number of communities, whereas for others it is the distribution of the community sizes. Second, and more importantly, it is possible to partition the network in order to get an estimated partition extremely similar to the real one, but whose misclassified nodes have a strong topological weight. For instance, misclassifying a hub will not change a partition-based measure much, but it can significantly affect the topological properties of the concerned communities. We conclude both approaches are complementary and needed to perform a relevant and complete analysis of community detection results. From a practical perspective, the traditional approach is much faster and easier to apply, so as a general guideline we propose to use it first. Then, the results corresponding to the best community structures can be more thoroughly inspected thanks to the topological measures.

Our contributions are as follow. First, we introduced a modification in the LFR model, in order to make the embeddedness distribution more realistic in the generated networks. Second, we studied these generated networks in terms of community-centred topological properties. This complements some previous analyses focusing on network-centred properties such as transitivity or degree correlation [7, 13, 14]. Third, we used these properties to compare community structures, by opposition to the traditionally applied performance-based approach [7, 8, 13, 14, 33]. Thanks to these tools, we were able to rank the tested community detection algorithms. Note, however, that these results might be specific to our benchmark, and we do not intent to generalize them to all kinds of networks. Our goal was rather to study the agreement between the various ranking methods.

This work can be extended mainly in two ways. First, the internal transitivity measured in the generated networks is very different from what is observed in real-world networks. The LFR model uses a random approach which does not favour the creation of triangles. According to our literature survey, there is no model generating networks with both a community structure and a controlled transitivity. So a solution would require either to define such a model, or to modify the LFR model again. Second, in this work we focused on a limited number of topological properties, classes of real-world networks, and community detection algorithms. A more thorough analysis would consist in expanding these factors. In particular, it would be interesting to include slower algorithms, which would allow comparing other types of community definitions. It would also be relevant to apply more classic network-wise measures to communities (degree correlation, centrality, etc.), and to consider additional community specific measures. Those designed in [34] seem particularly complementary to the embeddedness, and the concept of community profile [9] is promising, although it looks particularly costly from a computational point of view.